\documentclass[12pt]{article}
\newcommand{\nc}{\newcommand}

\nc{\dbar}{\bar{\partial}}
\nc{\be}{\begin{equation}}
\nc{\ee}{\end{equation}}

\usepackage{epsf,amsmath,latexsym}

\catcode`\@=11

\@addtoreset{equation}{section}
\def\theequation{\thesection\arabic{equation}}

\def\@normalsize{\@setsize\normalsize{15pt}\xiipt\@xiipt
\abovedisplayskip 14pt plus3pt minus3pt%
\belowdisplayskip \abovedisplayskip
\abovedisplayshortskip  \z@ plus3pt%
\belowdisplayshortskip  7pt plus3.5pt minus0pt}
\def\small{\@setsize\small{13.6pt}\xipt\@xipt
\abovedisplayskip 13pt plus3pt minus3pt%
\belowdisplayskip \abovedisplayskip
\abovedisplayshortskip  \z@ plus3pt%
\belowdisplayshortskip  7pt plus3.5pt minus0pt
\def\@listi{\parsep 4.5pt plus 2pt minus 1pt
            \itemsep \parsep
            \topsep 9pt plus 3pt minus 3pt}}

\def\underline#1{\relax\ifmmode\@@underline#1\else
        $\@@underline{\hbox{#1}}$\relax\fi}
\@twosidetrue
\relax

\catcode`@=12

\catcode`\@=11

\def\section{\@startsection{section}{1}{\z@}{3.5ex plus 1ex minus
   .2ex}{2.3ex plus .2ex}{\large\bf}}

\def\ps@headings{\def\@oddfoot{}\def\@evenfoot{}
\def\@oddhead{\hbox{}\hfill
        \makebox[.5\textwidth]{\raggedright\ignorespaces --\thepage{}--
        \hfill }}
\def\@evenhead{\@oddhead}
\def\subsectionmark##1{\markboth{##1}{}}
}

\ps@headings

\catcode`\@=12

\relax

\def\figcap{\section*{Figure Captions\markboth
        {FIGURECAPTIONS}{FIGURECAPTIONS}}\list
        {Fig. \arabic{enumi}:\hfill}{\settowidth\labelwidth{Fig. 999:}
        \leftmargin\labelwidth
        \advance\leftmargin\labelsep\usecounter{enumi}}}
 \relax
\def\tablecap{\section*{Table Captions\markboth
        {TABLECAPTIONS}{TABLECAPTIONS}}\list
        {Table \arabic{enumi}:\hfill}{\settowidth\labelwidth{Table 999:}
        \leftmargin\labelwidth
        \advance\leftmargin\labelsep\usecounter{enumi}}}
 \relax
\def\reflist{\section*{References\markboth
        {REFLIST}{REFLIST}}\list
        {[\arabic{enumi}]\hfill}{\settowidth\labelwidth{[999]}
        \leftmargin\labelwidth
        \advance\leftmargin\labelsep\usecounter{enumi}}}
 \relax

\catcode`\@=11

\def\ps@headings{\def\@oddfoot{}\def\@evenfoot{}
\def\@oddhead{\hbox{}\hfill
        \makebox[.5\textwidth]{\raggedright\ignorespaces --\thepage{}--
        \hfill }}
\def\@evenhead{\@oddhead}
\def\subsectionmark##1{\markboth{##1}{}}
}

\ps@headings

\relax

\def\firstpage#1#2#3#4#5#6{
\begin{document}

\begin{titlepage}
\nopagebreak
\title{\begin{flushright}
        \vspace*{-1.8in}
        {\normalsize hep-th/0203249}\\[4mm]
\end{flushright}
\vfill
{\large \bf #3}}
\author{\large #4 \\ #5}
\maketitle
\vskip -7mm
\nopagebreak
\begin{abstract}
{\noindent #6}
\end{abstract}
\vfill
\begin{flushleft}
\rule{16.1cm}{0.2mm}\\[-3mm]
March  2002
\end{flushleft}
\thispagestyle{empty}
\end{titlepage}}
\newcommand{\dal}{\raisebox{0.085cm}
{\fbox{\rule{0cm}{0.07cm}\,}}}
\newcommand{\dt}{\partial_{\langle T\rangle}}
\newcommand{\dtbar}{\partial_{\langle\bar{T}\rangle}}
\newcommand{\al}{\alpha^{\prime}}
\newcommand{\mst}{M_{\scriptscriptstyle \!S}}
\newcommand{\mpl}{M_{\scriptscriptstyle \!P}}
\newcommand{\dv}{\int{\rm d}^4x\sqrt{g}}
\newcommand{\lv}{\left\langle}
\newcommand{\rv}{\right\rangle}
\newcommand{\ph}{\varphi}
\newcommand{\sbar}{\,\bar{\! S}}
\newcommand{\xbar}{\,\bar{\! X}}
\newcommand{\fbar}{\,\bar{\! F}}
\newcommand{\zbar}{\,\bar{\! Z}}
\newcommand{\tbar}{\bar{T}}
\newcommand{\ubar}{\bar{U}}
\newcommand{\ybar}{\bar{Y}}
\newcommand{\phb}{\bar{\varphi}}
\newcommand{\cm}{Commun.\ Math.\ Phys.~}
\newcommand{\pr}{Phys.\ Rev.\ D~}
\newcommand{\pl}{Phys.\ Lett.\ B~}
\newcommand{\prl}{Phys.\ Rev.\ Lett.~ }
\newcommand{\ibar}{\bar{\imath}}
\newcommand{\jbar}{\bar{\jmath}}
\newcommand{\np}{Nucl.\ Phys.\ B~}
\newcommand{\e}{{\rm e}}
\newcommand{\gsi}{\,\raisebox{-0.13cm}{$\stackrel{\textstyle
>}{\textstyle\sim}$}\,}
\newcommand{\lsi}{\,\raisebox{-0.13cm}{$\stackrel{\textstyle
<}{\textstyle\sim}$}\,}
\date{}
\firstpage{95/XX}{3122} {\large\sc Open Strings on Plane waves and
their Yang-Mills duals} { David Berenstein$^a$, Edi Gava$^{b,c}$,
Juan Maldacena$^a$,\\
K.S.Narain$^c$ and Horatiu Nastase$^a$
 \,}
{\normalsize\sl
$^a$ Institute for Advanced Study, Princeton, NJ 08540 \\[-3mm]
\normalsize\sl
$^b$Istituto Nazionale di Fisica Nucleare, sez.\ di Trieste, SISSA,
Italy\\[-3mm]
\normalsize\sl $^c$The Abdus Salam International
Centre for Theoretical Physics,
I-34100 Trieste, Italy\\[-3mm]}
{We study the plane wave limit of $AdS_5\times S^5/Z_2$ which
arises as the near horizon geometry of D3-branes at an orientifold
7-plane in type I' theory. We analyze string theory in the
resulting plane wave background which contains open strings.
We identify gauge invariant operators in the
dual $Sp(N)$ gauge theory with unoriented closed and open string
states.
}

\section{Introduction}

Following the ideas in \cite{bmn} we consider a limit of a
particular large $N$ gauge theory where we can see both closed and
open strings arising in the large $N$ limit. These closed and open
strings move in a ten dimensional space, though the ends of the
open strings move on a D7 brane located at an orientifold
sevenplane. The limit is the same as the one
considered in \cite{bmn}. In \cite{bmn} only closed strings were
obtained since the Yang Mills theory contained only fields in the
adjoint representation. The theory we consider here is an ${\cal
N} =2$ $Sp(N)$ gauge theory with a hypermultiplet in the
antisymmetric representation  and four fundamental
hypermultiplets. Since there are fields in the fundamental we will
now have open strings in the 't Hooft limit \cite{tHooft}.
 This theory is dual
to string theory on $AdS_5\times S^5/Z_2$, where the $Z_2$ is an
orientifold action \cite{spalinski,ofer}.  Here we consider the
limit of large 't Hooft coupling, we pick a $U(1)$ generator, $J$,
inside the symmetry group of the theory and we consider operators
that carry large charge under $J$ but  have a conformal dimension
close to $J$, i.e. $\Delta -J$ is small. We consider such
operators at weak 't Hooft coupling and we identify some operators
that get small corrections to their dimensions which start looking
like strings propagating on an orientifold of the maximally
supersymmetric plane wave background \cite{blau}. We have closed
strings that arise from single trace gauge invariant operators as
in ${\cal N}=4$ SYM and open strings that arise from gauge
invariant operators with two quarks at the ends. We extrapolate
these results naively to strong coupling where they produce open
and closed strings moving in an orientifold background. The
orientifold projection is automatically obeyed by the gauge
invariant operators. The operators corresponding to open strings
are such that these open strings obey the appropriate boundary
conditions.

In section 2 we consider string theory in the background of the
orientifolded pp wave, in section 3 we look at the field theory
corresponding to the D3-D7-O(7) system. In section 4 we
identify the string theory
spectrum in the plane wave limit, with the gauge invariant operators.
We then analyze the calculation of
the anomalous dimension of operators, derive the string bit
hamiltonian and argue that we obtain the right boundary conditions
for the open string.

Other papers describing orbifolds of ${\cal N} =4$ gauge theories
and their corresponding plane wave limits include
\cite{klebanov,ooguri,zayas,jabbari,theisen,takayanagi,kehagias}.
Other aspects of plane waves were recently explored in 
\cite{russo,billo,cvetic1,gursoy,chu,cvetic2}.

\section{The orientifold of a plane wave }

We will consider the $O(7)$ orientifold projection with vanishing
tadpoles of the pp-wave solution of 10d IIB supergravity found in
\cite{blau}. An $O(7)$ plane carries $-4$ units of D-7-brane
charge, so  add 4 $D7$ branes to cancel the charge locally. This
produces a gauge group $SO(8)$ on the worldvolume of the
D-7-branes \cite{Polchinski}.

We want to take this orientifold of the pp-wave background. We
have directions $x^{\pm}$, and $y_{1,\dots 6}$ and the two extra
directions $y_{7,8}$. The orientifold acts by sending $y_{7,8}\to
-y_{7,8}$, and leaves all other coordinates fixed. This breaks
half of the 32 supersymmetries.

The metric is
\begin{equation}
ds^2= -4dx^+ dx^{-} - \mu^2 y^2(dx^+)^2 +  (dy^2) \label{muppwave}
\end{equation}
\begin{equation}
F_{+1234} = F_{+5678} = \mu
\end{equation}

The fixed point set of the orientifold is at  $y_{7,8}=0$. Notice
that the orientifold is parallel to the lightcone directions, so
that we can take  lightcone gauge and quantize the strings in this
background.  The quantization of the string theory on the pp wave
in light cone gauge was done in \cite{metsaev,mt}.
Once we choose
light cone gauge the orientifold projection will act as a
projection on the Hilbert space of the light cone gauge theory.

Let us first consider the closed string sector. Before the
projection the light cone action reduced to a set of eight massive
bosonic oscillators on a circle. We denote the corresponding creation and
annihilation operators by $a_{n}^{i \ \dagger} $,
$a_n^i$ where $n$ denotes the momentum of the oscillator along the
circle. Similarly we have eight massive fermions. All bosons and
fermions have the same mass. We denote the  fermionic annihilation and
creation operators by $b^a_{n}$ and $b^{a  \dagger}_n$ respectively.
The index
$a$ runs over the $(+,+)$ and $(- -)$ spinor representations of
$SO(4)\times SO(4)$ where $\pm$ denotes the chirality under each
$SO(4)$. The orientifold leaves the closed string  ground state
invariant and it
acts in the following way on the oscillators
\begin{equation}
a_n^i  \to  a_{-n}^i ~~~i=1,\dots 6  ~,~~~~~ a_n^{7,8} \to -
a_{-n}^{7,8} ~,~~~ ~~~
b_n  \to  i \Gamma^{56} b_{-n} \label{orientifoldaction}
\end{equation}
Note that on the Green Schwarz light cone fermions the orientifold
acts as $S_L \to  \Gamma^{78} S_R$, where $S_L$ and $S_R$
are real left and
right world sheet chirality fermions in definite chirality $SO(8)$ spinor
representation (Here and in the following $\Gamma_{78}$ for example denotes
$\Gamma_7 \Gamma_8$). However due to the fact that the
mass term is of the form $S_L \Gamma^{5678} S_R$ we find that on
the fermion creation operators $b_n^{a \dagger}$
 the orientifold action is
indeed as indicated in (\ref{orientifoldaction}).
So we see the usual statement that  the
orientifold reverses the direction of the string and it multiplies
the $y_{7,8}$ oscillators by $-1$.
If we only consider zero
momentum oscillators ($n=0$) then the orientifold projection
requires
that we have an even number of $a_0^{7,8  \dagger}$ oscillators.
We also have to remember to impose the $L_0-\bar L_0$ condition on
the states. The orientifold condition (\ref{orientifoldaction})
acting on the zero momentum fermions ($n=0$) multiplies four of
them by $-1$ and leaves the other four invariant. The four that
are left invariant are associated to supersymmetries that act
nonlinearly on the light cone gauge Hilbert space.

Now let us consider the open strings. For these strings, the
oscillators in the $y_{1\dots 6}$ directions have Neumann boundary
conditions, while the oscillators transverse to the orientifold,
$y^{7,8}$, have Dirichlet boundary conditions. The fermions obey a
boundary condition of the form  $ S_L =\Gamma^{78} S_R$. This
boundary condition leaves only 8 (real components) of the massive
fermion zero momentum modes $n=0$. These split into 4 creation and
4 annihilation operators (here by creation operators we mean
operators that increase the light cone energy). Again due to the
specific form of the mass term we find that:

i) all the creation operators carry charge $-1/2$ with respect to $SO(2)_{56}$
acting on $56$
directions, while all the annihilation operators carry charge $+1/2$.

ii) as a result, the ground state (i.e. lowest energy state) carries
$SO(2)_{56}$ charge $+1$.

iii) due to the unbalanced number of zero modes, four fermions and six
bosons, the open string ground state energy is $-p^-=\mu$ where $\mu$ is the
mass of the oscillators. All non-zero momentum oscillators
$(n\neq 0)$
come in equal numbers of bosons and fermions so that their contribution to
the ground state energy cancels.

These facts will be important for us
in the following when we identify these states with the gauge invariant
operators.

The action of these four creation operators generates the vector
supermultiplet of theories of 16 supercharges. As usual, the open
string ground state is odd under the projection \cite{Polchinski}
which  implies that the Chan-Paton  indices are anti-symmetric and
therefore are in the adjoint of $SO(8)$.  We can find all open
string states by keeping states invariant under the orientifold
projection. One way to think about it is to say that we can form
any state with the oscillators and then we arrange the SO(8)
indices into a symmetric or anti-symmetric representation to obey
the orientifold condition.

In appendix A, we give a more detailed proof of the above
statements.

\subsection{The orientifolded plane wave as a limit of
$AdS_5\times S^5/Z_2 $. }

We write the metric of $AdS_5\times S^5$ metric as \begin{equation}
ds^2=R^2(-\cosh^2\rho dt^2 +d\rho^2+\sinh^2\rho d\Omega_3^2 +
\cos^2 \theta d\psi^2 + d\theta^2 +\sin^2\theta d\Omega^2_3) 
\end{equation}
where the second $d\Omega_3^2$ on $S^3$ which is part of $S^5$ is
explicitly given as \begin{equation}
d\Omega_3^2=\cos^2\theta'd{\psi'}^2+d{\theta'}^2+\sin^2\theta' d\phi^2
\end{equation} 
The $Z_2$ action which combines with the orientifold
projection  is $\psi' \rightarrow \psi' +\pi$. In these
coordinates the D7 branes sit at $\theta'=\pi/2$.

We can now take the suitable limits to go to the pp wave solutions
exactly as in \cite{bmn}. This is a Penrose limit. The Penrose
limit consists in looking at the neighborhood of a lightlike
geodesic. If the geodesic does not intersect the orientifold plane
we get the maximally supersymmetric type IIB plane wave
\cite{blau}. It is more interesting to consider a geodesic that
lies on the orientifold plane, in this way we will retain the
orientifold plane as we take the limit. For this purpose we
consider the case where we boost along $\psi$.

We define \begin{equation}
 \rho = r/R,~~~~~\theta =y/R,~~~~~ x^+=\frac{(t+\psi)}
 {2} ,~~~~~x^- =  R^2\frac{t-\psi}{ 2} \label{coorddef} 
\end{equation} and
we take the limit $R\rightarrow \infty$. The metric then becomes
(\ref{muppwave}) with $\mu=1$. In these units we find that
 \begin{equation}
-p_- = {\Delta + J \over R^2} ~,~~~~~-p_+ = \Delta-J \end{equation}
where $\Delta$ is the energy conjugate to $t$ translations and $J$
is conjugate to translations in $\psi$. $-p_+$ is equal to the
light cone hamiltonian.

The $S_3$ (at $\theta'=\pi/2$) in $S^5$ where the orientifold is
lying has a symmetry $SO(4)=SU(2)_R \times SU(2)_L$, we can define
\begin{equation} \label{defofj}
J =  J^3_{SU(2)_R} + J^3_{SU(2)_L}~~,~~~J' =J^3_{SU(2)_R} -
J^3_{SU(2)_L}
\end{equation}

In terms of the coordinates in the metric (\ref{muppwave}) we see
that $J'$ performs rotations in the $y^{5,6}$ plane, while $J$ is
the one appering in (\ref{coorddef})

\section{ The gauge theory}

Consider a theory where we have $N$  $D3$ branes parallel to the
$O(7)$ plane. The $D3$ branes  have a gauge group $Sp(N)$. $N$
denotes the number of D3 branes after we do the orientifold (i.e.
in the covering space we have $2N$ D3 branes). We consider the low
energy limit where we decouple the   gauge theory living on the D3
branes from the rest. The $SO(8)$ gauge symmetry of the D7 branes
becomes a global symmetry of the theory on the D3 branes.

The theory on the $D3$ branes is an ${\cal N}=2$ theory, with a
hypermultiplet in the antisymmetric representation of $Sp( N)$.
These antisymmetric representations describe the motion  of the D3
brane along the directions of the D7 brane. We will split them
into two chiral fields $Z, Z'$. The scalar superpartner of the
gauge field will be called $W$, it describes motions of the branes
in the direction transverse to the orientifold. It can be shown
that the theory of the D3-branes at the origin in moduli space is
conformal \cite{lowe,bds,asyt}. We also have  the $D3-D7$ strings,
which are in the fundamental representation of $Sp(N)$. These are
four hypermultiplets in the fundamental, which we label in chiral
language by $q_i, \tilde q_i$. Since this representation is real
the global symmetry group is $SO(8)$, as expected also from the
fact that this is the gauge group of the D7 theory. This theory
has a $U(1)\times SU(2)_R$ R-symmetry and also a $SU(2)_L \times
SO(8)$ global symmetry. The chiral fields $Z, Z'$ are doublets of
$SU(2)_L$, they have zero charge under $U(1)$, and they, together
with their complex conjugates,   form a doublet of $SU(2)_R$.  The
field $W$ is  a singlet of both $SU(2)$ symmetries and carries
charge 1 under $U(1)$.   Finally the fields $q, \tilde q $ carry
zero $U(1)$ charge, are singlets under $SU(2)_L$ and together with
their complex conjugates they form a doublet of $SU(2)_R$.
Remembering the definition of the charge $J$ (\ref{defofj}) we
find that $Z$ has $J=1$ and $Z'$ has $J=0$. We also find that
$q^i, \tilde q^i$ have $J=1/2$.

We also have the relations
$-p_+ = \Delta - J$ and $-p_- \sim (\Delta + J)/R^2 $,
where $R^2$ is the curvature radius of $AdS$ in string units. In
summary, we consider the limit of large 't Hooft coupling and we
consider operators with large charge $J$ (with $J/R^2 \sim
J/(g^2_{YM}N)^{1/2} $ fixed), and small values of $\Delta-J$.

The superpotential of the theory is given by ${\cal N}=2$ SUSY. In
${\cal N} =1$ language it reads(up to normalizations of the
fields)
\begin{equation}
{\cal W} \sim (W_{ab} q_i^a \tilde q_i^b +W_{ab}
\Omega^{bc} Z_{cd} \Omega^{de} Z'_{ef}\Omega^{fa})
\end{equation}

The F-terms are given by
\begin{eqnarray} \label{fterms}
F_{\tilde q} = W q ~,~~~ F_{q} = W \tilde q \\
F_{Z'\ ad} =W_{ab}\Omega^{bc} Z_{cd} &-& (a\leftrightarrow d)\\
F_{Z \ ad}= W_{ab}\Omega^{bc} Z'_{cd} &-& (a\leftrightarrow d)\\
F_{W \ cb } = Z_{cd} \Omega^{da} Z '_{ab} &+& (b\leftrightarrow c)
+ (q\tilde q)_{(cb)}
\end{eqnarray}
The  action written in terms of components  has a potential which
is given by the square of the $F$ terms plus the square of the $D$
terms.

A chiral operator with angular momentum $J$  and $\Delta-J =0$ is
given by
\begin{equation}
(Z_{ab} \Omega^{bc})^J = {\rm tr} [(Z \Omega)^J]
\end{equation}
we need the invariant tensor $\Omega$ of $Sp(N)$ to raise one of
the indices so that we can use matrix multiplication. The indices
range from 1 to $2N$. This will be identified with the closed
string ground state with $-p_- =2J/R^2$.

There are other protected BPS operators, or
elements of the chiral ring, (see \cite{spalinski}) whose quantum
numbers are those of
\begin{equation}
{\rm tr} \left( (Z \Omega)^J(Z'\Omega)^{l_1} (W\Omega)^{2l_2}\right)
\end{equation}
More precisely, we have to symmetrize over the positions of $Z'
\Omega $ and of the $W \Omega$ insertions. The reason is that, as
usual, any operator proportional to $\partial {\cal W} =0$ (the
$F$ terms in (\ref{fterms})) is not a primary operator. The
$F_{Z}=0$ and $F_{Z '}=0$ equations tell us that $W \Omega$ can be
commuted freely past $Z \Omega$ and $Z ' \Omega $, and the
$F_{W}=0$ equation that $Z '\Omega $ can be commuted past $Z
\Omega$  \footnote{the commutator is a $q\tilde{q}$ which splits
the single trace into 2 traces, this is subleading in $1/N$ but it
seems to be important for interactions involving  open strings.}.
So the operator is actually
\begin{equation}
\sum _{i_1,...i_{l_1}; j_1,..., j_{l_2}=1}^J {\rm tr} ((Z
\Omega)^{i_1} (Z '\Omega) (Z \Omega)^{j_1-i_1}(W \Omega)....(W
\Omega)^ {j_{l_2}}(Z \Omega)^{J-j_{l_2}})
\end{equation}

Now $l_2$  should be even since  $Z  $ and $Z ' $  are
antisymmetric matrices, whereas $W $ is a symmetric matrix, thus
the symmetrized sum $\sum \left( (Z \Omega)^J(Z'\Omega)^{l_1}
(W\Omega)^s\right) $ is a symmetric matrix if s is even, and
therefore gives 0 when it multiplies another $W \Omega$. This
implies that we have an even number of $W$.

We can also have elements of the chiral ring with the same quantum
numbers as the operator
\begin{equation}
Q_i \Omega (Z \Omega)^{J} (Z'\Omega)^{l} Q_j
\end{equation}
where $Q_i$ $i=1,\cdots,8$  is any of $q^l, \tilde q^l$. The
F-terms imply that we should again symmetrize (sum) over all
possible positions  of the $Z' \Omega $ insertions and then the
operator
\begin{equation}
\sum_{k_1, ... k_l=1}^J Q_i \Omega (Z \Omega )^{k_1}(Z '
\Omega)... (Z '\Omega)^{k_l}(Z \Omega)^{J-k_l}Q_j
\end{equation}
is antisymmetric in $i,j$. These represent the massless open
strings at the orientifold fixed point \cite{ofer}

We cannot have elements of the chiral ring where we insert $W$ and
$Q$ simultaneously, because using the F-terms we can commute the
$W$ past anything else so that it lies adjacent to $Q$, and then
the operator vanishes when we impose that the $F_q$,
$F_{\tilde{q}}$  are zero. In other words, $F_{Z}=0$, $F_{Z '}=0$
tell us that we should sum over the positions of the $W \Omega$
insertions with equal weight, and $F_q=0$, $F_{\tilde{q}}=0$ tell
us that the weight at the endpoints should be zero \cite{ofer}.

\section{The string bit hamiltonian}

We now discuss the operators corresponding to the string states
that we found above.

The closed and open string ground states are identified as the
operators
\begin{equation}
{\rm tr} ( (Z\Omega)^J) \label{groundstateclosed}
\end{equation}
\begin{equation}
Q_i \Omega(Z\Omega)^J Q_j \label{groundstateopen} \end{equation}
Note that $\Delta -J =0$ for  (\ref{groundstateclosed}), while
$\Delta - J=1 $ for  (\ref{groundstateopen}) since $\Delta -J
=1/2$ for $Q_i$. This is in precise  agreement with the
non-vanishing ground state energy that we found for open strings
above. Furthermore, since $Q$ is $SU(2)_R$ doublet, it follows that
the operator (\ref{groundstateopen}) carries $J'=+1$, which agrees
with what we found for the open string vacuum. 
Replacing the operators $Q_j\leftrightarrow Q_i$ does not
produce a new state, but just a minus sign. This follows  because
$\Omega(Z\Omega)^J$ is antisymmetric, and so the $i,j$ indices are
in the adjoint of SO(8). Of course this had to work since it is a
rephrasing of the comparison done in \cite{ofer}.

In analogy with the discussion in \cite{bmn}  we associate each
oscillator on the string with the insertion of an operator with
$\Delta -J=1$  along the ``string of $\Omega Z$s'' that we have in
(\ref{groundstateclosed}, \ref{groundstateopen}). The first four
$y^i$ oscillators are associated to the insertion of a derivative,
as in \cite{bmn}. $y^{5,6}$ are related to the insertion of $Z'$
or $\bar {Z'}$.  The $y^{7,8}$ oscillators are related to the
insertion of $W$ and $\bar W$  in the operator. We can also insert
the fermion operators with $\Delta-J=1$. We have the fermions in
the vector multiplet and also the fermions in the antisymmetric
representation which are in a hypermultiplet. Once we have chosen
the components with $J=1/2$ these two types of fermions are
distinguished by their $J'$ eigenvalue. On the string Hilbert
space these two correspond to the two possible eigenvalues of
$i\Gamma^{56}$ on $b^{a\dagger}_n$. Then the string states are
obtained by summing over all possible positions inside the trace
of the insertions of $\partial_i Z, Z ', \bar Z ', W, \bar W$ and
fermions with $J=1/2$. We can act with these on either the closed
string ground state, or the open string ground state. We have seen
that if we only act with zero momentum modes (corresponding to
elements of the chiral ring), we can only act with an even number
of $a_0^{7,8 \dagger}$'s (corresponding to $W, \bar W$ insertions)
on the closed string ground state, and with no $a_0^{7,8 \dagger}
$'s on the open string ground state. Similarly the fermion zero mode
creation operators 
in open string come with $J'=-1/2$ which translates into the gauge
theory statement that the chiral ring involves only the insertion of
anti-symmetric fermion and not the adjoint ones. That is as it should be, from
the orientifold projection. Once we include nonzero momentum
oscillators, we can act with any number of $a_{n}^{7,8
\dagger}$'s.

To construct nonzero modes we need to add momentum on the
worldsheet, and the momentum $n$ is related to a phase
proportional to the $J$ charge to the left of the operator, as in
\cite{bmn}. Let us see  how the orientifold projection arises for
them. Any state that does not obey the orientifold projection is
automatically zero once we take into account the symmetry or
anti-symmetry of the various indices of the operators. The action
of the orientifold is essentially taking the transpose of the
operator. For example,  we can consider the operator
\begin{equation}
\label{operator}
\sum_l  e^{i 2 \pi l n \over J}
Tr\left[W \Omega (Z \Omega)^l Z' \Omega (Z \Omega)^{J-l} \right]
\end{equation}

We can think of this state as  $a_{-n}^{7+i8 \dagger} a_{n}^{5+i6
\dagger} | 0,p_-\rangle_{l.c.}$.   Under the orientifold action
this state is mapped to $ -a_{n}^{7+i8 \dagger}
a_{-n}^{5+i6\dagger} | 0,p_-\rangle_{l.c.} $. So the  surviving
state is the combination $(a_{-n}^{7+i8 \dagger} a_{n}^{5+i6
\dagger} -a_{n}^{7+i8 \dagger} a_{-n}^{5+i6\dagger} )
|0,p_-\rangle_{l.c.}$. Now we would like to prove that the
operator (\ref{operator}) can be interpreted as containing both
terms in precisely this combination. To see this we can take the
transpose of (\ref{operator}). We get a minus sign from  the
transpose of $W $.  And we easily see that we effectively change
the momentum from $n \to -n$.  So we see that we indeed have the
precise combination we have  in the lightcone after performing the
orientifold. The two possible $J=1/2$ fermions, the one coming
from the vector multiplet and the one from the antisymmetric
hypermultiplet, get a relative minus sign when we take the
transpose.
 As we said above these
two fermions are distinguished by their eigenvalue of $J'$. This
is related to the fact that under the orientifold projection the
fermions on the string worldsheet get a sign proportional to their
$J'$ eigenvalue.

The $L_0 - \bar L_0$ condition is imposed by the cyclicity of the
trace as in \cite{bmn}.

The large N Feynman diagrams of this theory are similar to the
SU(N) theory except that the double lines that represent particles
do not have oriented edges. They have unoriented edges.
This
implies that that we can build non-orientable surfaces, etc. When
we twist a field in the symmetric or anti-symmetric
representations we get different minus signs \cite{cicuta}.

In our case we will be interested only in planar diagrams. So the
only difference with SU(N) will be in  the labeling of states, as
we discussed above. The gauge invariant states automatically
implement the orientifold projection.

The derivation of the string hamiltonian is completely parallel to
what we discussed in \cite{bmn}.  Here we just note that there is
a simple way of doing the computation if we keep track of what
terms in the bosonic potential come from the F terms and which
from the D terms.

\begin{figure}[hbt]
\centerline{ \epsfxsize 1.9in \epsfbox{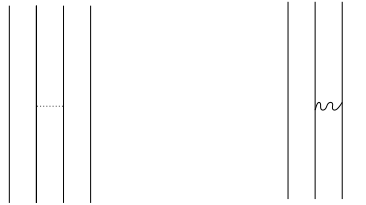} } \caption[]{The
corrections involving D terms, photons and self-energies cancel
each other if all fields are holomorphic. They do not exchange the
position of the fields. We denoted the propagator of the auxiliary 
$D$ field by a doted line, this is  just a delta function in  position 
space.   } \label{figdterms}
\end{figure}

Then it becomes rather simple to compute the conformal dimension
of an operator of the form
\begin{equation}
\sum_l  e^{i 2 \pi nl \over J}  Tr[ \cdots Z \Omega  (Z' \Omega)
(Z \Omega)^{J-l} ]
\end{equation}
which contains only chiral fields but with an extra phase for the
field $Z'$.  These are holomorphic operators. Then we know that
when the phases are zero,  all diagrams that contribute to the
anomalous dimension vanish. In this case diagrams coming from the
$F$ terms cancel by themselves, while the ones coming from
photons, D-terms  and self energy corrections of each individual
propagator,  cancel each other (see figure \ref{figdterms} ).
These latter diagrams do not change the position of the fields,
so they will also cancel when we include phases.

\begin{figure}[hbt]
\centerline{ \epsfxsize .3in \epsfbox{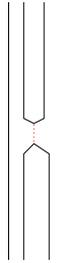} } \caption[]{The
diagrams involving F terms give non vanishing contributions which
involve terms exchanging the position of the fields. 
 We denoted the propagator of the auxiliary 
$F$ field by a doted line, this is  just a delta function in  position 
space. }
\label{figfterms}
\end{figure}

So the total anomalous dimension comes from the diagrams involving
$F$ terms (see figure \ref{figfterms}). These diagrams exchange
the position of two fields and give a result that depends on the
phases. Since the $F$ terms contain commutators we see that the
diagrams in figure \ref{figfterms} correctly contain the relative
$(-1)$ signs between the terms that exchange position relative to
the ones that do not exchange position. This is why the contributions
to the anomalous dimensions gives terms that involve $(Z'_i
-Z'_{i+1})^2$ in the effective action of the string.

Now let us discuss the open strings in more detail. The bulk of
the open string is the same as the bulk of the closed string. New
aspects arise when an excitation aproaches the boundary.  We need
to understand the origin of the Dirichlet and Neuman boundary
conditions. In principle, when we insert excitations with some
phases we can put phases that mimic any boundary condition. So we
need to understand the first order correction to the Hamiltonian.
If we insert a field $Z'$ at position $k$ along the ``open string
of $Z$'' in (\ref{groundstateopen}) we find an effective
Hamiltonian of the form
\begin{equation}
\sum_k  g^2_{YM} N (b_k+b_k^\dagger - (b_{k-1}-b_{k-1}^\dagger))^2
+ b_kb^\dagger_k
\end{equation}
We are interested in the term proportional to $g^2N$. This is a
quadratic form which is proportional to the following matrix
\begin{equation} M =
\begin{pmatrix}
1 &-1&  0 &0 &\dots\\
-1& 2 &-1& 0& \dots\\
0& -1 &2 &-1 &\dots\\
\vdots &\ddots  &\ddots& \ddots &\vdots\\
0&\dots&0&-1&1
\end{pmatrix}
\end{equation}
and it is a discretized version of the second derivative. Here the
only thing to notice are the terms at the upper left and bottom
right corner of the matrices which come from examining the
potential terms coming from $F$ terms in (\ref{fterms}). We want
to write the theory in terms of normal modes, so we want to write
it in terms of eigenvectors of the above matrix. It is a standard
exercise in coupled harmonic oscillators to show that the
eigenvectors of the above matrix have the form $\cos{ \pi n k/J}$.
So that we effectively have Neumann boundary conditions. One can
also see this explicitly by looking at the first component of the
eigenvalue problem for $M$, $(M-\lambda) v=0$. This takes the form
$v^1 - v^2 = \lambda v^1$, since the eigenvalue will be
proportional to $1/J^2$ in the limit of large $J$ we see that this
reduces to $ \partial_\sigma v \sim v(0)/J \sim 0$ in the large
$J$ limit.

Similarly we can consider an insertion of a $W$ field along the
``open string of $Z$s'' (\ref{groundstateopen}). Now the matrix to
be diagonalized has the form
\begin{equation}
W =
\begin{pmatrix}
2 &-1&  0 &0 &\dots\\
-1& 2 &-1& 0& \dots\\
0& -1 &2 &-1 &\dots\\
\vdots &\ddots  &\ddots& \ddots &\vdots\\
0&\dots&0&-1&2
\end{pmatrix}
\end{equation}
where the only difference with the previous one is the factor of
$2$ in the lower right and upper left corners of $W$. This arises
because there is an extra potential term coming from the $|F_q|^2$
terms in the bosonic potential. Again it can be seen that we
recover the Dirichlet boundary condition. One can see this by
looking at the first equation for an eigenvector $v^k$ of
eigenvalue $\lambda$. This becomes $ v(0) - \partial_\sigma v/J
\sim \lambda v(0) $ which becomes $v(0) =0$ in the $J\to \infty$
limit.

In summary, we see that open strings obey the appropriate boundary
conditions. These boundary conditions depend on the form  of the
interactions in the gauge theory.

\section{Conclusions}

In this paper we have considered the dual pair of ${\cal N}=2$
superconformal $Sp(N)$ gauge theory and the string theory in the
near horizon geometry  $AdS_5\times S^5/Z_2$ in the Penrose limit.
In particular for the Penrose limit which includes the D7 branes,
string theory has open string sectors. Following the proposal of
\cite{bmn} we compare a class of gauge invariant operators in the
gauge theory with the oscillator states of the string theory, both
in the unoriented closed and open string sectors. In particular,
we find that the operators which are bilinear in the fundamental
hypermultiplets are dual to open string states. By studying the
eigenvectors of the string bit hamiltonian we find the Dirichlet
and Neumann boundary conditions at the level of gauge invariant
operators.

A natural extension of this paper is to consider open strings ending on
baryons. The BPS sector of those open strings was considered in
\cite{bhk}. If we consider a baryon or a giant graviton with $J=N$
which are represented by D3 branes on $S^5$ and we take the
Penrose limit along a geodesic on the D3 then we find a plane wave
with a D3 along $x^\pm$ and two of the transverse coordinates. In
this case we expect that the open string spectrum is given again
by putting oscillators with phases on the open strings discussed
in \cite{bhk}. We plan to study this further.


{\bf Acknowledgements}

This  research was supported in part by DOE grants DE-FGO2-91ER40654,
DE-FG02-90ER40542 and by the EEC contract SCI-CT920792

{\large\bf{Appendix A: String spectrum on the orientifold of a
plane wave}}
\renewcommand{\theequation}{A.\arabic{equation}}
\setcounter{equation}{0}

In this appendix we discuss in more detail some aspects of the
string spectrum on the orientifolded plane wave.

{\large\bf{Closed strings}}

Let us recall
the basic facts. In the light cone gauge $x^+=p^+\tau$, the 8
transverse bosonic and Green-Schwarz fermionic coordinates become
massive. The mode expansion for the light cone variables are
\cite{mt}:
\begin{eqnarray}
x^I(\sigma, \tau)&=&
i\sum_{n}\frac{1}{\sqrt{2\omega_n}}(e^{-i\omega_n\tau+in\sigma}a^I_n-
e^{i\omega_n\tau-in\sigma}(a^I_n)^{\dagger})\nonumber\\
\theta^1(\sigma,\tau)&=& [e^{-i\mu p^+ \tau} b_0 -\sum_{n>0}c_n
e^{-i\omega_n
\tau}(e^{in\sigma}
b_n +\frac{\omega_n- n}{\mu p^+}e^{-in\sigma}
b_{-n})]+c.c.
\nonumber\\
\theta^2(\sigma,\tau)&=&[-e^{-i\mu p^+ \tau} i \Pi b_0 -i\Pi \sum_{n>0}c_n
e^{-i \omega_n\tau}(e^{-in\sigma}
b_{-n}-\frac{\omega_n- n}{\mu p^+}e^{in\sigma}
b_{n})]+c.c.
\label{mode}
\end{eqnarray}
where \begin{equation}
 \omega_n=\sqrt{n^2 +(\mu p^+)^2}, \end{equation}
 $c_n$ are some
normalization constants which we shall not need here and $\Pi$ is
the reflection operator  $\Gamma_{5678}$ along 4 directions
$x^5,..,x^8$. This can be compared with eqs.(2.5-2.7) of
\cite{mt}, by the substitution $(a^I_n)^{\dagger}=
\sqrt{\frac{\omega_n}{2}} \alpha^{1I}_{n}$ and
$(a^I_{-n})^{\dagger}= \sqrt{\frac{\omega_n}{2}} \alpha^{2I}_{n}$
for $n>0$. The creation and annihilation operators are
$a_n^{\dagger}$ and $a_n$ respectively for all integers $n$.
Similarly $b_n = \theta^1_{n}$, $b_{-n}=i\Pi \theta^2_{n}$ for
$n>0$ and the zero mode $b_0 = \theta^1_0+i\Pi \theta^2_0$. For
all integers $n$, $b_n$ and $b_n^{\dagger}$ are annihilation and
creation operators respectively. Note that this choice of vacuum
for the fermion zero modes
corresponds to the lowest energy state and the light cone Hamiltonian
comes with vanishing zero point energy. The level matching
condition in this notation becomes the vanishing of the sum of the
momenta $n$ of  all the oscillator modes of the creation
operators, $\sum_i n_i=0$. This form of level matching condition
is convenient for the comparison with the gauge theory operators.

The $Z_2$ projection in the present case is $\Omega (-1)^{F_L} R$
where $R$ reflects the two directions $\vec{y}$. On the fermions
$R$ therefore acts as $\theta^I \rightarrow \Gamma_{78} \theta^I$.
The world sheet parity operator exchanges $\theta^1$ with
$\theta^2$ while $(-1)^{F_L}$ gives a minus sign to $\theta^1$.
Combining all this one finds that the $Z_2$ action on the
oscillator modes is
\begin{eqnarray}
a_n^I &\rightarrow& a_{-n}^I,~{\rm for}~ I=1,..,6; ~~~~
a_n^I \rightarrow -a_{-n}^I,~{\rm for}~ I=7,8; \nonumber \\
b_n &\rightarrow& i\Gamma_{56}b_{-n}
\end{eqnarray}

Note that, in particular only half of the fermion zero modes
are $Z_2$ even, signalling that supersymmetry is now half
compared to the type IIB case. In particular we note that these
$Z_2$ even fermion zero mode annihilation (creation) operators  come
with definite eigenvalue w.r.t.
$i\Gamma_{56}$ implying that they have $J'$ charge $-1/2$ ($+1/2$). This is
important in the identification with the gauge theory chiral operators
since this $J'$ charge corresponds to an insertion of the anti-symmetric
fermion as opposed to the adjoint fermion.

The physical states are the ones that are even under the above
$Z_2$ and satisfy level matching condition. We take the vacuum
state $|0,p^+>$ to be even under $Z_2$. The states that are
obtained by applying bosonic creation operators are
\begin{equation}
\psi(n_1,..,n_k; I_1,..,I_k)= \Pi_{r=1}^k (a_{n_r}^{I_r})^{\dagger}
|0,p^+>
\label{psi}
\end{equation}
which under $Z_2$ action transform as
\begin{equation}
\psi(n_1,..,n_k, I_1,..,I_k) \rightarrow
(-1)^{\ell}\psi(-n_1,..,-n_k; I_1,..,I_k)
\label{psi1}
\end{equation}
where $\ell$ is the number of creation operators in $\psi$ along 7 and 8
directions.

{\large{\bf{Open string sector}}}

In type I', there are also open strings that are stretched between
the 8 D7 branes. In the Penrose limit under consideration these D7
branes are located at the origin of the 78 plane. This means that
$x^7$ and $x^8$  must satisfy Dirichlet boundary conditions. The
boundary conditions at
$\sigma=0$ and $\pi$ are therefore:
\begin{equation}
\partial_{\sigma}x^I =0,~~{\rm for}~ I=1,..,6; ~~~~~~ x^I=0 ~~{\rm for}~
I=7,8~~~~~~~ \theta^1= \Gamma_{78}\theta^2
\end{equation}
The general
solutions to the equations of motion in the light-cone gauge
subject to the above boundary conditions are:
\begin{eqnarray}
x^I(\sigma, \tau)&=& i\sum_{n \geq
0}\frac{1}{\sqrt{2\omega_n}}\cos n\sigma (e^{-i\omega_n\tau}
a^{I}_n-e^{i\omega_n\tau} (a^{I}_n)^{\dagger}) ~~{\rm
for}~I=1,..,6
\nonumber\\
x^I(\sigma, \tau)&=& i\sum_{n > 0}\frac{1}{\sqrt{2\omega_n}}\sin
n\sigma (e^{-i\omega_n\tau} a^{I}_n-e^{i\omega_n\tau}
(a^{I}_n)^{\dagger}) ~~{\rm for}~I=7,8
\nonumber\\
\end{eqnarray}
The solution for the $\theta^1$ and $\theta^2$ are the same as in
the closed string case eq.(\ref{mode}) subject to the condition
\begin{equation}
b_n = i\Gamma_{56}b_{-n}
\label{bc}
\end{equation}
There are a few important points to note. While the
bosonic Neumann directions $x^I$ for $I=1,..,6$ have zero modes,
the two Dirichlet directions $x^7$ and $x^8$ do not have zero
modes.  For fermions there are on the other hand 8 zero modes (as
opposed to 16 in the closed string case) of which 4 are creation and 4
annihilation operators. From (\ref{bc}) it follows that all the annihilation
operators $b_0$ carry the $SO(2)_{56}$ charge $J'=+1/2$. Since the state
$(b_0^{\dagger})^4 |0>$ carries $+2$ units of $J'$ charge above that
of the vacuum $|0>$ we conclude that $|0>$ must carry $J'=+1$.
The vacuum energy now is $\mu$ due to the fact that
the number of boson zero modes is 6 while that of fermion zero modes
is 8. We will see this fact also by matching the states obtained by
applying the fermion zero mode creation operators on the vacuum state
to the vector multiplet living on the D7 brane world volume.

We can decompose the states in the vector multiplet of D7 brane
in the light cone gauge in terms of $SO(4)\times SO(2)_{56}\times
SO(2)_{78}$ where $SO(4)$ is the rotation group acting on $1234$
directions,
(recall that $J'$ is the $SO(2)_{56}$ charge). The vector multiplet
contains a light cone vector which has the 4 components $V_i$
transforming
as $(v,0,0)$,  and a complex $V$ (and $\bar{V}$) in $(1,\pm 1,0$). The
complex scalar $\phi$ (and $\bar{\phi}$)
in the vector multiplet transforms as $(1,0,\pm)$. The fermions are
$\psi_{\pm}$ transforming as $(sp,\pm1/2, \mp 1/2)$ and $\psi'_{\pm}$
in $(sp',\pm 1/2, \pm 1/2)$. Here $v$, $sp$ and $sp'$ denote the vector,
spinor and spinor'
representations of $SO(4)$.  Denoting by $b_{0L}$ and $b_{0R}$ the 
projections $(1+\Pi)b_0/\sqrt{2}$ and $(1-\Pi)b_0/\sqrt{2}$,
we get the above table for the
bosonic states.
\begin{table}
\begin{tabular}{lccc}
{\bf open string states}~~~~~&$E_0/\mu$~~~~~~~&$SO(4)\times SO(2)_{56}
\times SO(2)_{78}$ ~~~~~~~&{\bf vector multiplet states}\\
$|0>$~&c~~~~~~&$(1,+1,0)$ ~~~~&V\\
$(b_{0L}^{\dagger})^2|0>$~~~~&c+2~~~&$(1,0,-1)$ ~~~~~&$\bar{\phi}$\\
$(b_{0R}^{\dagger})^2|0>$~~~~&c+2~~~~~&$(1,0,+1)$~~~~~&$\phi$\\
$b_{0L}^{\dagger}b_{0R}^\dagger |0>$~~~~&c+2~~~~&$(v,0,0)$~~~~&$V_i$\\
$(b_{0}^{\dagger})^4|0>$~~~~&c+4~~~~~&$(1,-1,0)$~~~~&$\bar{V}$
\end{tabular}
\end{table}
In this table 
$c$ is the energy of the open string vacuum. To show that
$c=1$ we need only to determine its value
from the scalar field equation for
say $\phi$.
Recalling that D7 branes are located at the origin of the 78
plane, we can carry out the analysis of the linearized equations
of motion for these fields in the pp wave background exactly as in
\cite{mt}. For example the scalar field equation in the pp
background $\Box \phi =0$  gives rise to the following light cone
Hamiltonian:
\begin{equation}
H=-p_+ = \frac{1}{2p^+}\sum_{I=1}^6 (p_I^2 -\mu^2
{p^+}^2 \partial_{p_I}^2)
\end{equation}
This is just the Schrodinger
equation for a non-relativistic 6-dimensional harmonic oscillator
and the zero point energy $E_0$ is simply $6/2 \mu =3\mu$.
Comparing with the table above and noting that the scalar fields
appear in 2nd and 3rd rows in the table, we conclude $c=1$. A similar
analysis
for gauge fields can be carried out where the Chern-Simons terms
play a crucial role in splitting the energies of $V$, $\bar{V}$ and $V_i$.

Finally,  the $Z_2$ action on  these
different oscillator modes are as follows (as can be seen by  taking $\sigma$
to $\pi-\sigma$ together with $(-1)^{F_L}$ and the  reflection in the 78
plane):   \begin{equation}
a^I_n \rightarrow (-1)^n
a^I_n ,~~~I=1,\cdots 8~;
 ~~~~~~~
b_n \rightarrow (-1)^n  b_{n}
\end{equation}

\end{document}